%% file: imgrelations.tex
\documentclass[conference]{acmsiggraph} %,draft

\pdfoutput=1

\usepackage{amssymb}
\usepackage{bm}

\TOGonlineid{0221}
\TOGvolume{0}
\TOGnumber{0}
\TOGarticleDOI{1111111.2222222}
\TOGprojectURL{}
\TOGvideoURL{}
\TOGdataURL{}
\TOGcodeURL{}

\usepackage{xspace} 

\newcommand{\raid}{\textsc{raid}\xspace}

\title{RAID: A Relation-Augmented Image Descriptor}

\author{Paul Guerrero\thanks{paul.guerrero@ucl.ac.uk}\\{KAUST, University College London} \and Niloy J. Mitra\thanks{n.mitra@cs.ucl.ac.uk}\\University College London \and Peter Wonka\thanks{pwonka@gmail.com}\\KAUST}
\pdfauthor{Paul Guerrero, Niloy J. Mitra, Peter Wonka}


\begin{document}

\teaser{
  \vspace{-10pt}
  \includegraphics[height=1.5in]{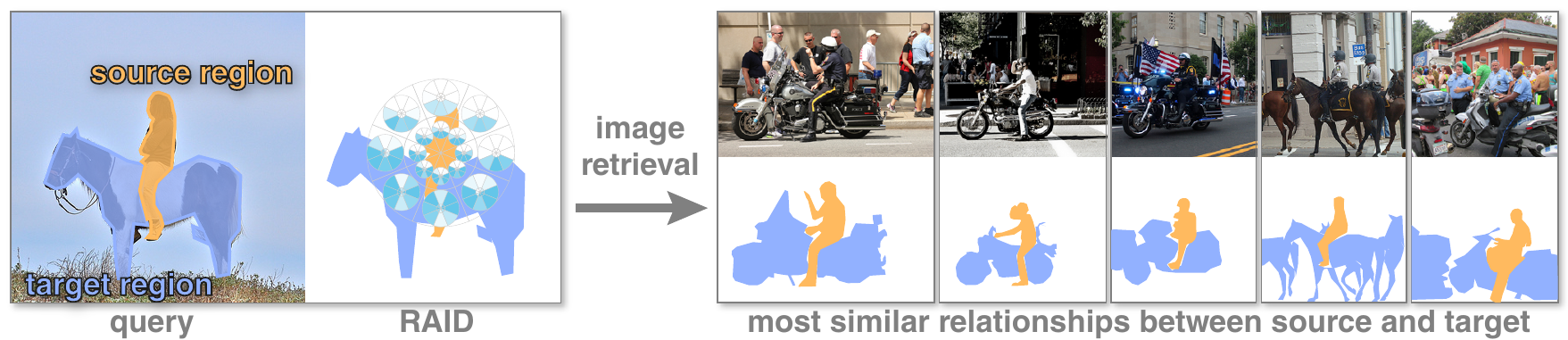}
  \caption{We propose a novel descriptor called \raid to describe the spatial relationship between image regions. This descriptor enables image retrieval with queries based on complex relationships between regions, such as the `riding' relationship between the orange source and the blue target region.}
  \label{fig:teaser}
  \vspace{-5pt}
}

\maketitle

\input{abstract.tex}

\if0
\begin{CRcatlist}
  \CRcat{I.4.8}{Scene Analysis}{Shape};
  \CRcat{I.4.7}{Feature Measurement}{Size and Shape};
%  \CRcat{I.3.7}{Computer Graphics}{Three-Dimensional Graphics and Realism}{Radiosity};
\end{CRcatlist}
\fi

\keywordlist

%\TOGlinkslist

%\copyrightspace

\input{introduction.tex}

\begin{figure*}[t]
	\includegraphics[width=1.0\textwidth]{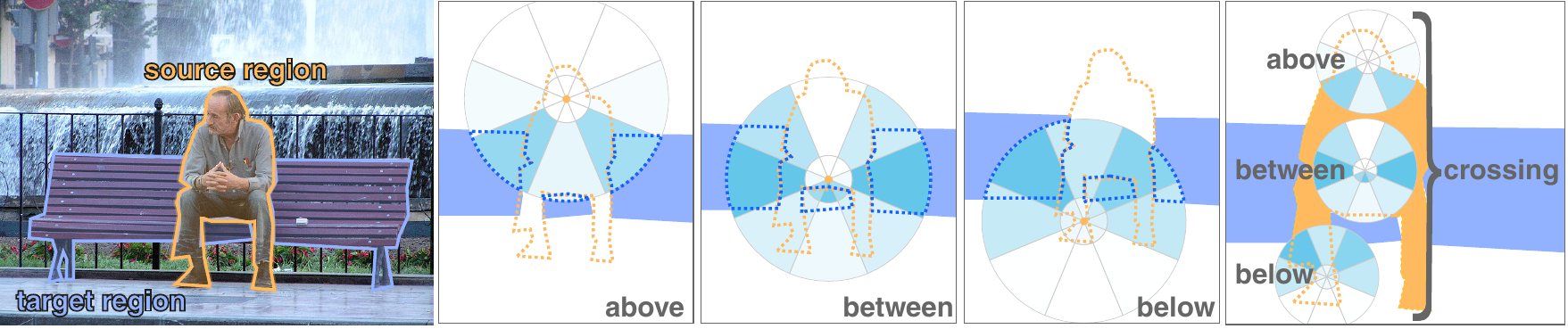}
	\caption{\footnotesize Simple and complex relationships between the man and the bench shown on the left. %Given the `man' source region and the `bench' target region,
		We can identify several simple relationships between points in the source region (man) and the target region (bench). The relationships of each point is described by a polar histogram, with each bin colored according to the percentage of overlap with the target region. Some points are above the bench, some are below, and some are in between the bench segments. When looking at the spatial distribution of these simple relationships, we can infer the more complex `crossing' relationship between source and target region.}
	\label{fig:descriptoronreal}
	\vspace{-10pt}
\end{figure*}

\section{Related Work}

%There has been some research into 

%The idea of using spatial relationships between image regions to enhance content-based image retrieval has been 
%However, most methods focus on the 

Most research on spatial relationships between image regions has been done in the field of content-based image retrieval. These methods usually focus on describing the composition of \emph{all} regions in an image and use relatively simple models for individual pair-wise relationships. The survey of Bloch~\shortcite{Bloch2005} gives a good overview of early methods that include statistics over distances or directions (although not both) between points in both regions. In these methods, no attempt is made to describe complex relationships or capture a spatial distribution of relationships.
More recent approaches can be classified by the type of models they employ: 
%that are employed to describe region relationships:

% early work
%most notable use of region relationships in content-based image retrieval
%Histogram of orientations, 
%see ... for a good overview of methods published before 2005 (only fuzzy?)
%- \cite{Bloch2005} - STAR on spatial relationships - methods only have 1D angle histograms over all pairs of points or 1D distance histograms with different distance measures, not both combined  (See Page 98, first paragraph and Section 6.3). Complex relationships are not considered (surrounded is the most complex relationship considered, which we still classify as a simple relationship).
%several image retrieval methods use relationship models based on the spatial relationships of region centroids
\paragraph{Point-based relationships models.} One class of methods
%~\cite{Ko2002,Lee2002,Lan2012,Huang2014}
represent each image region as a single point, usually the centroid or bounding box center. As a consequence, only simple relationships such as the distance~\cite{Ko2002} or the direction~\cite{Lee2002,Lan2012,Huang2014} between the representative points are captured (including relationships like `below' and `above'). A richer description of region relationships is presented in Zhou et al.~\shortcite{Zhou2001}, based on the directional interval subtended by one region relative to the centroid of the other region. Complex relationships between two regions, however, can not be captured since one of the regions is still represented as point.

%`extent' of one region relative to the centroid of the other region

%In these methods, each region are represented by a single point, usually the centroid or the center of the bounding box. Ko et al.~\shortcite{Ko2002} use the distance between region centroids 

%- \cite{Ko2002}, a predefined \emph{spatial location point}~\cite{Lee2002}, the geometric centers~\cite{Huang2014}
%- \cite{Ko2002} - relationship is distance between region centroids, Hausdorff-like distance measure to match two sets of regions (also provide overview of relationship-based image retrieval methods published before 2000)
%- \cite{Lee2002} - Image objects represented as points, graph with simple relationships between points (below, above, etc.) is constructed.
%- \cite{Lan2012} - Only above, below and overlap relationships (relations seem to be between bounding box centers).
%- \cite{Huang2014} - Fully connected graph of superpixels, spatial relationships are directions between geometric centers of superpixels.
%- \cite{Zhou2001} - Directional 'extent' of object B as seen from centroid of object A (see Figure 3 in paper).

%a good overview of methods based on histograms of angles, 

%string-based relationship representations for image retrieval
\paragraph{String-based relationships models.} A different line of research uses strings to describe the spatial layout of regions in an image~\cite{Wang2003,Hsieh2008}. These methods project the image regions to the x- and y-axes of the image and record the starting point and end point of each projected region in two strings: one for the x-axis and one for the y-axis. This provides a compact representation of the region layout. However, a lot of information is lost during the projection to the image axes, resulting in a less discriminative description of relationships (for example, `surrounded' cannot be distinguished from `in a concave').
%- \cite{Wang2003} - Be-String: Works on Bounding Rectangles of objects. Describes spatial relations as two strings: one for x-axis, one for y-axis. Each String contains the start- and endpoints of each bounding rectangle in order from min. coordinate value to max. coordinate value.
%=> Only 1D projections of bounding boxes, can't capture complex relationships, e.g. riding where bounding boxes may be practically identical.
%- \cite{Hsieh2008} - Regions are represented by symbols and they are treated as points for the purpose of describing spatial relationships (see Figure 2 of paper).

\paragraph{Adjacency-based relationship models.} Several methods \cite{Chandran2003,Badadapure2013} describe the layout of image regions as a graph, where nodes correspond to regions and edges connect adjacent regions. Region layouts can be compared efficiently using techniques from graph theory. Again, no attempt is made to describe complex relationships or the spatial distribution of relationships over a region. Similar to our paper, Hu et al.~\shortcite{Hu2013} try to find matching regions in a large image library based on inter-region relationships. Relationships between adjacent image regions are described by a histogram of the relative locations between border pixels in a small 2-pixel neighborhood. This allows capturing simple relationships between adjacent regions like `above' or `below'. In contrast, our approach describes a spatial distribution of relationships, enabling us to capture more complex relationships between image regions that do not need to be adjacent.
%This approach is limited to adjacent regions and since the spatial distribution of relationships is not captured, only simple relationships like `above' or `below' can be described.

%However, only adjacent regions are considered

%also store the  of adjacent image regions in a small 2-pixel neighborhood. This allows capturing simple relationships between adjacent regions like `above' or `below'.
%Even though

%Again, 

%for content-based image retrieval, content described as adjacency graph between image regions
%- \cite{Chandran2003} - Regions are described as graphs with edges for adjacent regions. Graph matching is used to compare images. => Can't handle complex relationships, only adjacency.
%- \cite{Badadapure2013} - Based on only containment (containment trees) and adjacency (segmentation graph) of regions.
%- \cite{Hu2013} - PatchNet: only 5x5 pixel histogram of relative pixel positions between two regions (see Section 3.1 of paper). Retrieves patches, not images based on the PatchNet graph
%due to the limited size of the window, only adjacencies can be considered and the authors do not attempt to capture the spatial distribution of these adjacencies.

\paragraph{Scene understanding.} An important part of scene understanding is to accurately identify the relationship between scene objects. Several methods tackle this challenge by creating models of region relationships. Malisiewicz and Efros~\shortcite{Malisiewicz2009} encode the spatial context of image regions in a graph. Features used in the spatial context are the amount of overlap, relative displacement, relative scale and relative height between two regions. Kulkarni et al.~\shortcite{Kulkarni2013} use one specialized detector for each of their $16$ simple relationship classes like `above', `on,', and `near'. Adding an additional class requires implementing an additional detector. Recently, Karpathy et al.~\shortcite{Karpathy2015} presented a deep learning method to create natural language descriptions of images, with impressive results when trained on large datasets. However, it does not have an explicit representation of relationships.
Our approach describes more complex relationships, provides a single data-driven descriptor for all relationship classes and does not need to be trained on a large dataset.

%- \cite{Karpathy2015} - Learn image labels and relationships labels jointly using a neural net. For describing images, not for queries. Also a lot of training is necessary for the neural net to work properly, we only require a single pair of regions to do a query, no training is necessary.
%- \cite{Kulkarni2013} - Only simple relationships like above, below, near. Each computed with a very simple detector specialized to that one relationship (e.g. A above B is the percentage of the bounding box of object A above the top of the bounding box of object B).
%- \cite{Malisiewicz2009} - Relationships used:
%relative overlap, relative displacement, relative scale, and relative height of the bottom-most pixel (Section 3.2 of paper)
%=> no complex relationships

\paragraph{Shape descriptors.} Several shape descriptors have been proposed over the last two decades. Surveys can be found in \cite{Zhang:2004,Kazmi:2013}. Some region-based shape descriptors can be adapted to describe the simple relationship between a point and an image region. These include polar and square shape matrices~\cite{Goshtasby:1985,Flusser:1992}, moment-based shape descriptors~\cite{Teague:80,Celebi:2005} and Shape Contexts~\cite{Belongie:2002}. In this work we describe a novel descriptor for complex relationship between two image regions. We use Shape Contexts~\cite{Belongie:2002} as a baseline shape descriptor to compare the performance of our method.

%------ Maybe not necessary:

%- \cite{Ahmad2003} - Method works on feature points of the image objects: 2D points computed by any type of feature extraction (e.g. corners). Organize the features in a quadtree, comparet two quadtrees to compare two images.
%=> Features are unreliable, might not always be available and are not descriptive enough. The quadtree can only capture absolute positions in an image, not relationships between feature points in the same image.
% => quadtree captures absolute position of image features, does not have so much to do with relation between image regions

%- \cite{Zhang2013} - Not very related, distance between the region centroid and the expected position for a region of the same label is used to enhance image labeling. (e.g. sky is expected to be at the top of the image, so a sky label of the image will have more weight if the sky region is at the top).

%- \cite{Lee2012} - Not very related, above and below relationships based on region centroids.
%(They build histograms of posterior probabilities that a segment with a given label appears above or below a central segment.)

\section{Relationships Between Image Regions}

\begin{figure*}[t]
	\includegraphics[width=1.0\textwidth]{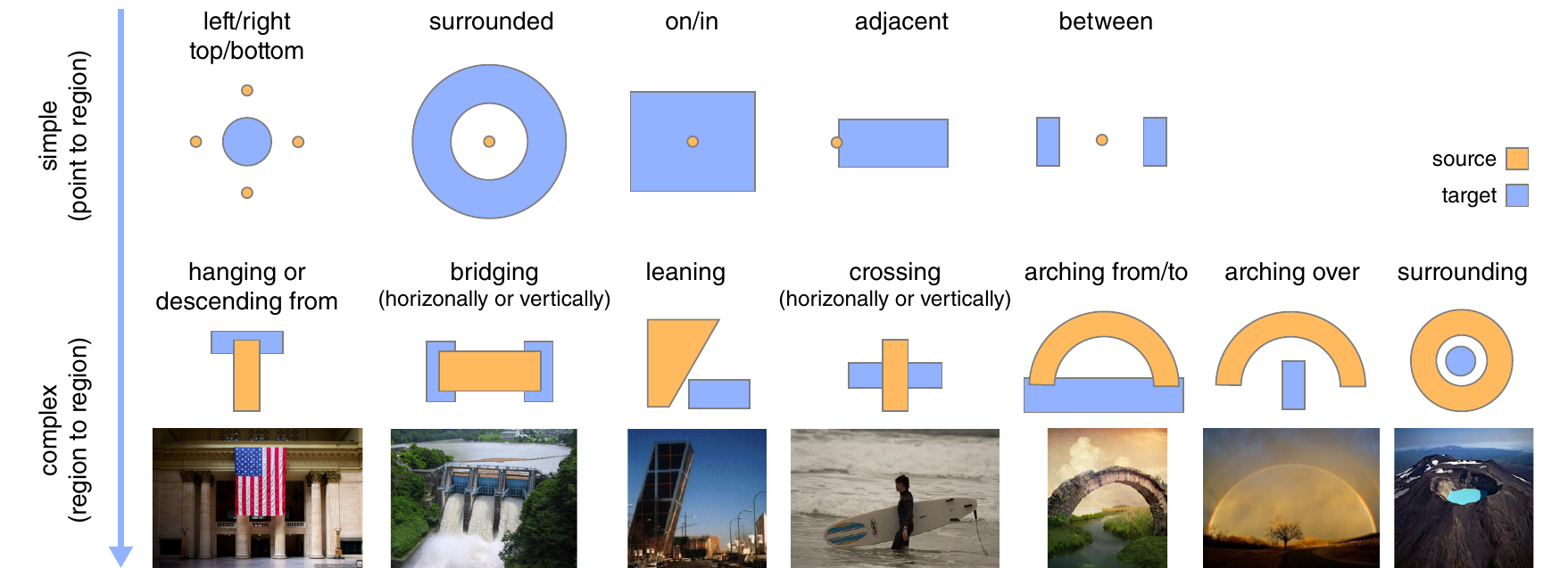}
	\caption{\footnotesize Classes of spatial relationships between two-dimensional image regions. We distinguish simple relationships (top row) and complex relationships (bottom row) between the orange source region and the blue target regions. Example images are shown below each complex relationship.}
	\label{fig:relationships}
	\vspace{-12pt}
\end{figure*}

Here, we provide a definition of spatial relationships between two image regions and give several examples of such relationships.
While most images that we consider are two-dimensional projections of three-dimensional scenes, our goal is to describe the two-dimensional composition of image regions rather than inferring a three-dimensional layout of the scene and then analyzing relationships in three dimensions. The advantage of this design choice is that the approach is a lot more robust, because inferring three-dimensional layouts from a single image is a challenging and underdetermined problem.

We can identify several classes of relationships that are commonly encountered in images. Examples are `between', `bridging', `arching', `crossing', as shown in Figure~\ref{fig:relationships}. We can observe, that most of the relationships are asymmetrical. For example, if a region $A$ is to the left of region $B$, then region $B$ is to the right of region $A$. We therefore need to distinguish between the two regions involved in a relationship and we call the first region in a relationship the source region, and the second region the target region.

For the purpose of this paper, we use a simple categorization to distinguish between simple and complex relationships. A simple relationship is one that exists for source points as well as source regions. For example, both a point and a region can be surrounded by another region. A complex relationship can only exist for source regions larger than a single point. For example, only a region and not a point can surround another region. Hence, the `surrounded' relationship is simple and the `surrounding' relationship complex. In Figure~\ref{fig:relationships} examples of simple relationships are shown on top and examples of complex relationship are shown on the bottom.

While there are several well-established methods to describe simple relationships, most importantly Shape Contexts~\cite{Belongie:2002}, in this paper we set out to design a descriptor to describe complex relationships as well as simple ones.

We use the following definitions:

The domain $I$ of an image is a rectangular subset of $\mathbb{R}^2$. An image region $A$ is defined as a subset of $I$. A labeling of an image region is a function $l:\mathbf{A} \rightarrow L$ where $\mathbf{A}$ is the set of all image regions and $L$ is a label set. %Typically, we are interested in regions in $I$ that have the same label.

A relationship class is a function that assigns a binary class membership to a pair of regions:
\begin{equation*}
C_x(S, T) = 
\begin{cases}
1 & \text{if}\ S\ \text{is in relationship}\ x\ \text{with}\ T \\
0 & \text{otherwise.}
\end{cases}
\end{equation*}
Note that the same pair of regions can be members of multiple relationship classes. Further, in some datasets, labeled regions are disjunct (e.g. the COCO dataset) while some other data sets allow for overlaps between labeled regions (e.g. the synthetic and web datasets).
In the next section, we propose a novel descriptor that is able to encode complex relationships.

\section{The RAID Descriptor}

\begin{figure*}[t]
	\includegraphics[width=1.0\textwidth]{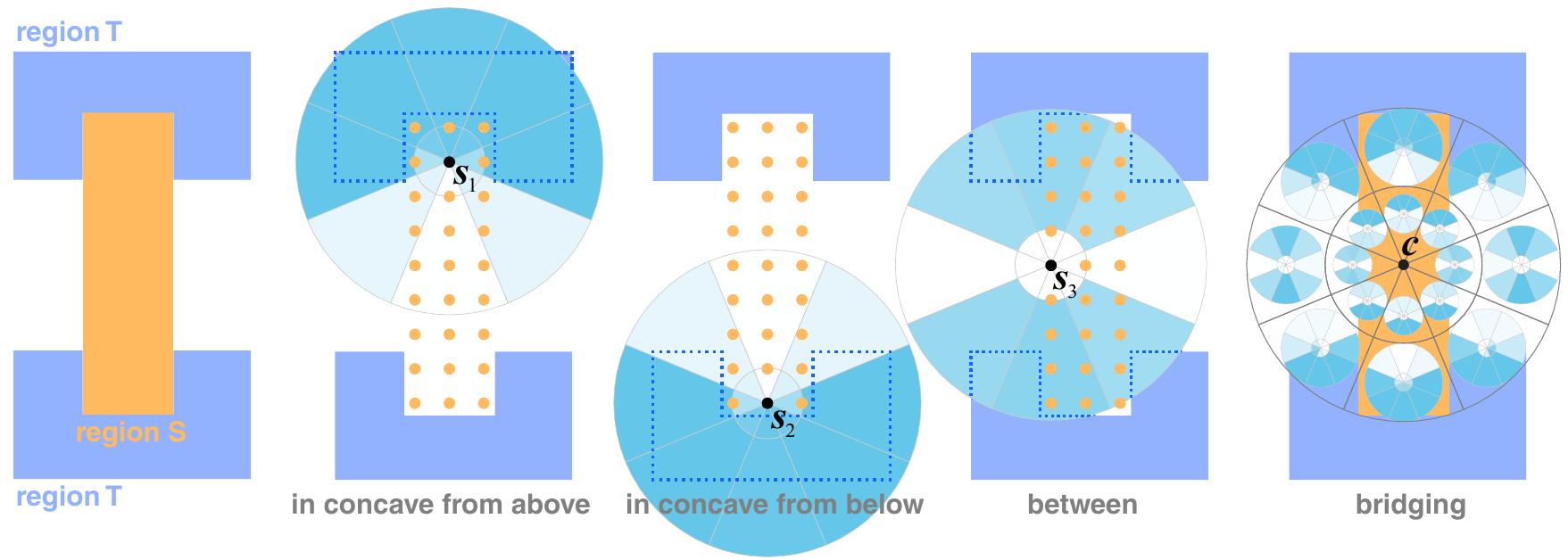}
	\caption{\footnotesize The \raid descriptor of the relationship between two image regions $S$ and $T$.
		%of paths starting at the centroid $p$ of region $A$, continuing to a point $q$ in $A$ and ending at a point $r$ in $B$.
		In this example region $S$ `bridges' region $T$ vertically.
		%The last segment of this path describes simple geometric relationship between individual points $q$ in $A$ and region $B$. he fir
		Simple relationships between individual points $\mathbf{s}$ in $S$ and the region $T$ are described by histograms of relative distance and direction from $\mathbf{s}$ to points in $T$: $\mathbf{s}_1$ and $\mathbf{s}_2$ are in a concave of $T$, while $\mathbf{s}_3$ is between $T$.
		%of each point $q$ in A to region B are described with a histogram of the relative distance and direction to points $r$ in B.
		More complex relationships between regions $S$ and $T$ are characterized by the distribution of simple relationships over $S$, which we capture in a histogram of simple relationships (rightmost image). In the `bridging' relationship shown here, points like $\mathbf{s}_3$ that are between $T$ are added to bins closer to the centroid $\mathbf{c}$, while points like $\mathbf{s}_1$ and $\mathbf{s}_2$ that are in a concave part of $T$ contribute to bins further above and below. Note that the histograms in each bin on the right are scaled down for illustration only; they have the same size as the histograms shown in the center images.}
	\label{fig:descriptor}
	\vspace{-12pt}
\end{figure*}

The aim of our descriptor is to provide a numerical description of the relationship between a given source region $S$ and a given target region $T$. We build on the fundamental observation that a complex relationship between $S$ and $T$ can be characterized by the relationship of each point in $S$ to each point in $T$.
Our approach to build the descriptor was therefore to first describe the relationship of each point in $S$ to the region $T$ separately. Afterwards, the problem becomes finding a suitable way to aggregate all the individual point to region descriptors. In the following we will describe our solution to encode the distribution of point relationships over $S$.

A point relationship is described by a two-dimensional histogram $H(\mathbf{s})$ of distance and direction between a source point $\mathbf{s}$ and each point $\mathbf{t}$ in the target region, similar to Shape Contexts~\cite{Belongie:2002}:
\begin{eqnarray}
H_{ij}(\mathbf{s}) = \frac{1}{a_{ij}}\int_{\Phi_i} \int_{R_j} \mathbf{1}_{T}(\mathbf{s} + (r \cos \phi, r \sin \phi)^\mathrm{T})\ r\ dr\ d\phi,
%H_{ij}(\mathbf{s}) = \frac{\int_{\Phi_i} \int_{R_j} \mathbf{1}_{T}(\mathbf{s} + (r \cos \phi, r \sin \phi)^\mathrm{T})\ r\ dr\ d\phi}{\int_{\Phi_i} \int_{R_j} 1\ r\ dr\ d\phi},
%H^{\mathbf{s}}_{i,j} = \int_{\mathbb{R}} \mathbf{1}_{\mathbf{T}}\ \mathbf{1}_{\mathbf{B_{i,j}}}\ d\mathbf{p} \\
%\mathbf{B_{i,j}} = \{\mathbf{p}\ |\ atan2(\mathbf{p}) \in \Phi_i, ||\mathbf{p}|| \in R_j\},
\label{eq:pointhistogram}
\end{eqnarray}
where $\Phi_i$ and $R_j$ are the angular and radial intervals of bin $(i,j)$, and $\mathbf{1}$ is the indicator function. Each bin is normailzed by the bin area $a_{ij}$. We call this histogram the \emph{point histogram}. Figure~\ref{fig:descriptor}, center shows an example for two regions in the bridging relationship for points $\mathbf{s}_1$, $\mathbf{s}_2$ and $\mathbf{s}_3$. Basically, a histogram bin will contain a value corresponding to the fraction of its area covered by region $T$.
%of $1$ if its complete area is covered by region $T$ and $0$ if $T$ does no intersect the histogram bin.

The distribution of point relationships over the source region is then encoded by a second histogram $\mathcal{H}^{S}$ over the individual point histograms, resulting in a four-dimensional histogram:
\begin{eqnarray}
%\mathcal{H}^{\mathbf{S}}_{i,j,k,l} = \int_{\mathbf{B}} \int_{R_k} H^{\mathbf{s}}_{i,j} \mathbf{1}_{\mathbf{S}}\ d\mathbf{p},
\hat{\mathcal{H}}^{S}_{ijkl} = \frac{\int_{\Phi_k} \int_{R_l} (\mathbf{1}_{S} H_{ij})(\mathbf{c} + (r \cos \phi, r \sin \phi)^\mathrm{T})\ r\ dr\ d\phi}{\int_{\Phi_k} \int_{R_l} \mathbf{1}_{S}(\mathbf{c} + (r \cos \phi, r \sin \phi)^\mathrm{T})\ r\ dr\ d\phi},
\label{eq:raiddescriptor}
\end{eqnarray}
where $\mathbf{c}$ is the centroid of the source region. Figure~\ref{fig:descriptor}, right shows an illustration of the 4D histogram. The denominator normalizes each bin by the intersection of the bin area with the source region. This factors out the dependence of the histogram on the exact shape of the source region and only captures the distribution of point histograms. Bins with zero intersection are assigned the value of the point histogram at the closest point of the source region. Finally, we perform a histogram normalization:
\begin{eqnarray}
%\mathcal{H}^{\mathbf{S}}_{i,j,k,l} = \int_{\mathbf{B}} \int_{R_k} H^{\mathbf{s}}_{i,j} \mathbf{1}_{\mathbf{S}}\ d\mathbf{p},
\mathcal{H}^{S}_{ijkl} = \frac{\hat{\mathcal{H}}^{S}_{ijkl}}{\sum_{ijkl} \hat{\mathcal{H}}^{S}_{ijkl}}.
\label{eq:raidhistogramnorm}
\end{eqnarray}
We call this histogram the \raid descriptor. Similar to the SIFT descriptor~\cite{Lowe04}, the \raid descriptor is a histogram of histograms, but \raid encodes directions and distances to a target region while SIFT encodes gradient orientations.

\section{Implementation}

In our implementation, we assume that image regions are given as polygons.
The integral for the point histogram in Equation~\ref{eq:pointhistogram} can then be computed accurately and efficiently by constructing the Boolean intersection between the target region polygons and a set of polygons representing each bin of the point histograms. A performant and robust implementation of this operation is available in the Boost polygon library~\cite{Boostpolygon}.

The integral in Equation~\ref{eq:raiddescriptor} involves finding a point histogram for each source point. An analytical solution
%would involve intersections of each bin with complex four-dimensional geometry, which would result an inefficient and unnecessarily complex method.
%Therefore,
is not feasible, therefore,
we resort to an approximation. First, point histograms are computed at a regular grid of samples $\mathbf{s}$ inside the source region. As a good tradeoff between performance and accuracy, the density is chosen to be approximately $10000 / a_I$, where $a_I$ is the image area. Due to the limited sample density, directly accumulating these point histograms in the bins of the \raid descriptor would result in considerable aliasing, especially for smaller bins. Instead, we approximate the integral over a bin with a sum over all samples, weighted by a Gaussian kernel centered inside the bin:
\begin{eqnarray}
%\hat{\bm{\mathcal{H}}}^{S}_{ijkl} = \frac{\sum_{\mathbf{s} \in \{\mathbf{s}\}} \mathbf{H}^\mathbf{s}_{ij} \mathcal{G}(\mathbf{s} | \mathbf{c} + (r_j \cos \phi_i, r_j \sin \phi_i)^\mathrm{T},\sigma^2)\ r\ dr\ d\phi}{\int_{\Phi_i} \int_{R_k} \mathbf{1}_{S}(\mathbf{c} + (r \cos \phi, r \sin \phi)^\mathrm{T})\ r\ dr\ d\phi},
\hat{\bm{\mathcal{H}}}^{S}_{ijkl} = \frac{\sum_{\mathbf{s} \in \mathbf{S}} H_{ij}(\mathbf{s}) \mathcal{G}(\mathbf{s} | \mathbf{c} + \mathbf{b}_{kl},\sigma^2)}{\sum_{\mathbf{s} \in \mathbf{S}} \mathcal{G}(\mathbf{s} | \mathbf{c} + \mathbf{b}_{kl},\sigma^2)},
\end{eqnarray}
where $\mathbf{S}$ is the set of samples inside the source region, $\mathbf{b}_{kl}$ is the centroid of bin $(k,l)$ and $\mathcal{G}(\mathbf{x}|\bm{\mu},\sigma^2)$ is an isotropic two-dimensional Gaussian with mean $\bm{\mu}$ and variance $\sigma^2$.
The variance of the Gaussian is chosen so that the volume under the function equals the volume under the characteristic function of the bin. Note that this is a relatively coarse approximation, but it is efficient and works well as long as the shape of the bins is not too thin and elongated. As in Equation~\ref{eq:raidhistogramnorm}, the final discretized descriptor is then obtained through histogram normalization:
\begin{eqnarray}
\bm{\mathcal{H}}^{S}_{ijkl} = \frac{\hat{\bm{\mathcal{H}}}^{S}_{ijkl}}{\sum_{ijkl} \hat{\bm{\mathcal{H}}}^{S}_{ijkl}}.
\end{eqnarray}

In all our experiments, we set the maximum distance $r_{\mathrm{max}}$ for the outermost bin in the \raid descriptor to the maximum distance between source region centroid and any other point in the source region. This ensures that the \raid descriptor covers the entire source region and effectively makes the descriptor scale-invariant. The maximum distance for the point histograms is set to the same value, meaning that an offset of $r_{\mathrm{max}}$ around the source region is captured by our descriptor.  Our implementation uses 8 bins for both angular dimensions and 2 bins for both radial dimensions, giving a total of 256 bins. The descriptor geometry is shown in Figure~\ref{fig:descriptor}. Center images show the size of bins $(i,j)$ relative to the source region, the rightmost image shows the size of bins $(k,l)$ (note that the histograms shown inside each bin $(k,l)$ are scaled down for illustration only). Rotational invariance could be achieved by aligning the descriptor to the first principal component of the points in the source region. However, on many types of images, rotational invariance is not desirable (e.g. `bridging horizontally' is different from `bridging vertically'), therefore we keep the descriptor aligned to the x-axis of the image.

%radius of the point descriptors  as well as the to half the span of the source region (the distance between the two farthest points).

\begin{figure*}[t]
	\includegraphics[width=1.0\textwidth]{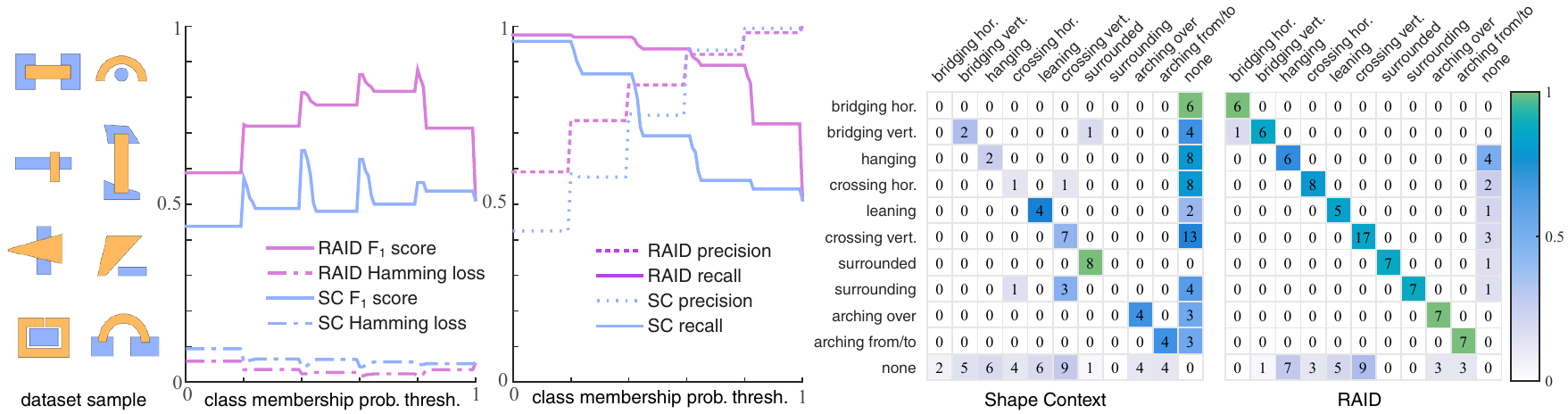}
	\caption{\footnotesize Classification performance on the synthetic dataset and comparison to Shape Contexts. On the left, we show part of the dataset, followed by various performance measures at different class membership probability thresholds of the binary k-NN classifiers. On the right, the confusion matrices for Shape Contexts and \raid are shown (rows correspond to actual classes, columns to predicted classes, colors are normalized by class size, while numbers show absolute values). Note that Shape Contexts generally perform worse and are unable to detect some relationship classes, like `bridging' or `surrounding'.}
	\label{fig:artificaleval}
	\vspace{-12pt}
\end{figure*}

\section{Evaluation and Applications}

%- Querying databases by example (artificial examples or non-artificial examples, evaluate precision for artificial and large non-artificial datasets)

%- Querying with a sentence (?) (using pre-defined relationship templates, evaluate similar to querying by example)

%- Classification of relationships (evaluate precision/recall on artificial and small non-artificial dataset)

To evaluate the performance of our descriptor, we performed experiments on 10000 images of the Microsoft COCO dataset~\mbox{\cite{Lin2014}}, a smaller synthetic dataset, and a small dataset of images collected from the web. The COCO subset contains a large variety of photographs that are suitable to evaluate the real-world performance of our method. However due to its large size, annotating every relationship to measure classification performance is not feasible. Instead, we perform image retrieval on this dataset and annotate the $n$ best results of each query. This ground truth is used to evaluate the precision of our method. The synthetic dataset contains several abstract shapes and is small enough to exhaustively annotate all relationships. We evaluate precision as well as recall on this dataset. To measure classification performance on real images, we could take a small subsample of the COCO dataset. However, this would result in severe undersampling of the more uncommon relationship classes. Considering this, we collected a set of $69$ images from the web instead, containing a balanced mix of relationship classes. All datasets were finalized before starting our experiments.

To the best of our knowledge, currently there exists no descriptor that explicitly attempts to describe complex relationships between image regions. Most methods only describe simple relationships, that is, relationships that can also be found between a point and a region. In the following evaluations, we compare our method to Shape Contexts~\cite{Belongie:2002}. Since our descriptor uses histograms similar to Shape Contexts to describe simple relationships, this comparison also demonstrates how adding information about the distribution of simple relationships results in a description that is better suited for complex relationships.

\paragraph{Computational Complexity and Performance}
Computing our descriptor has a complexity of $O(N_s N_b)$, where $N_s$ is the number of sample points in the source region and $N_b$ the number of bins of the point histogram. Since the number of bins is constant, the complexity is linear in the area of the source region.
Our simple single-threaded Matlab implementation requires approximately $0.13$ seconds per descriptor on average. The COCO subset contains roughly $236000$ relationships (24 relationships per image on average), which gives a total time of $8.5$ hours for an exhaustive query on the entire dataset. However, specifying a label for the source or target region lowers the number of relationships by a factor of typically $4$--$5$. Additionally, we can precompute the descriptors for the entire dataset, which requires about $510$ MB of space. Querying the dataset then only requires computing the $L_1$ distances between the query descriptor feature vector and the feature vectors of the precomputed descriptors, which requires roughly $0.46$ seconds in our Matlab implementation.

\begin{figure*}[p]
	\includegraphics[width=1.0\textwidth]{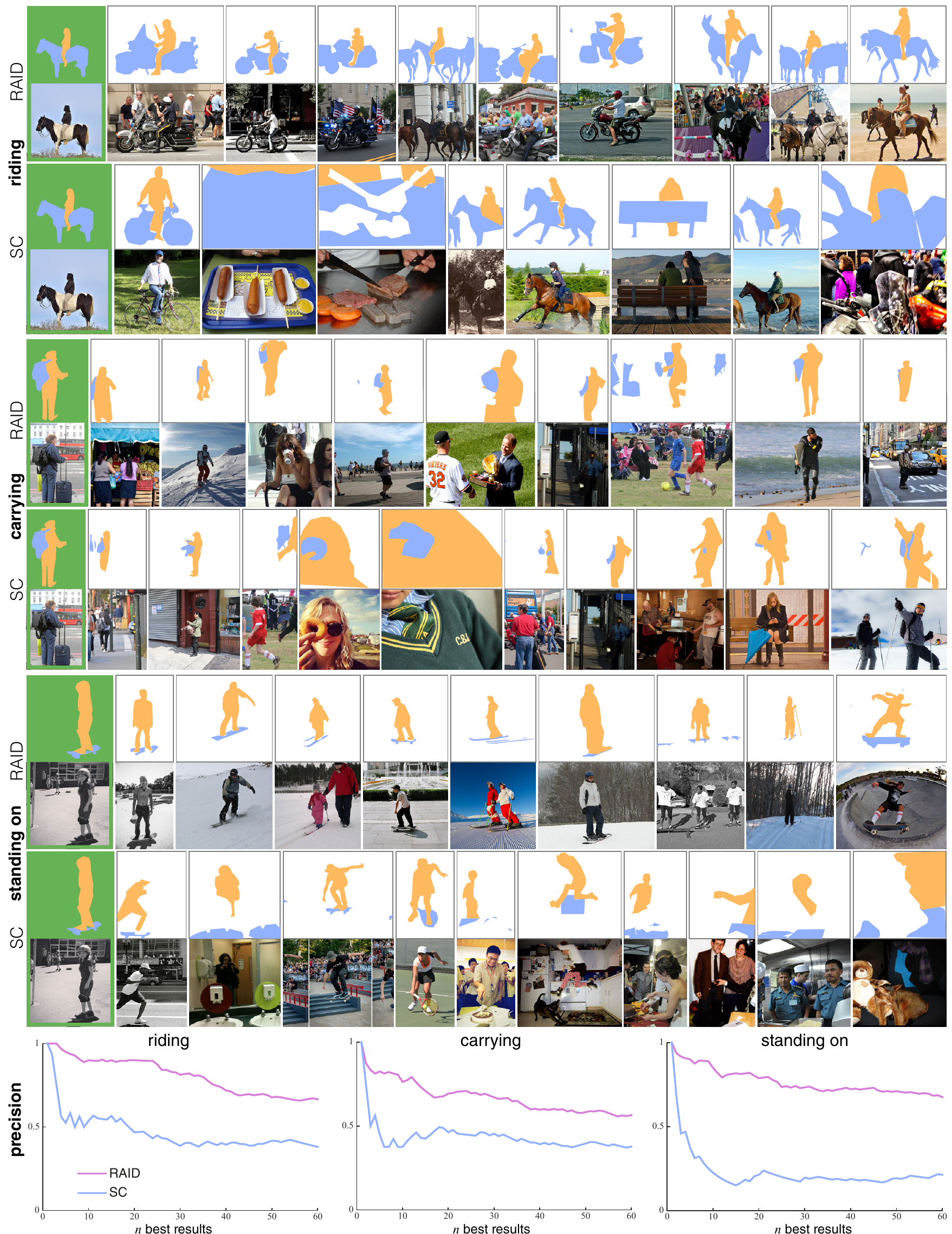}
	\caption{\footnotesize Three relationship queries between the orange source regions and the blue target regions shown in the first column (green background). Source regions are set to be persons, while target regions may have any label. Results are shown for the \raid descriptor and Shape Contexts (SC). In each row, we show the $n$ best results for the query shown in the first column. The bottom row shows the precision of the $n$ best results as a function of $n$. Note how the \raid descriptor finds regions that are intuitively more similar to the query relationship.}
	\label{fig:queryresultsA}
\end{figure*}

\begin{figure*}[p]
	\includegraphics[width=1.0\textwidth]{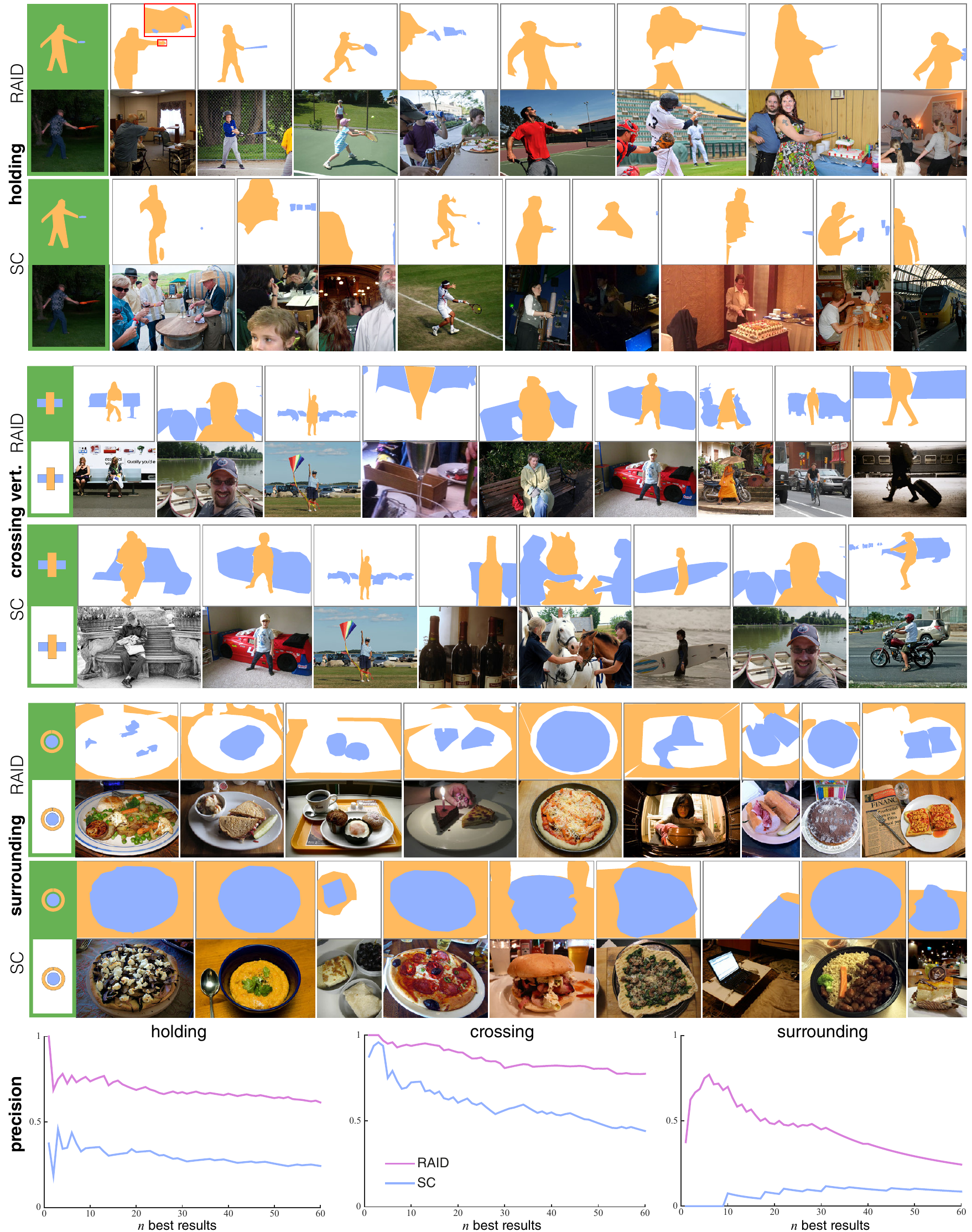}
	\caption{\footnotesize Three additional relationship queries between the orange source regions and the blue target regions shown in the first column (green background). In the first query, source regions are set to be persons, while target regions may have any label. The second and third query were specified with a synthetic source and target region and relationships with any source and target labels were searched.}
	\label{fig:queryresultsB}
\end{figure*}

\paragraph{Image Retrieval} An interesting area of application for the \raid descriptor is image retrieval from large databases. Our method can extend the search capability of a system by enabling queries for given relationships, such as `riding' or `standing on'. In the following we present experiments we have performed with different relationship queries on a dataset of 10000 images from the Microsoft COCO dataset~\cite{Lin2014}. In this dataset, image regions are annotated by labeled polygons. The set of labels is consistent throughout the dataset and the annotation quality is relatively high, which makes it a good choice for our experiments.

To specify a relationship query, we can either mark a pair of regions in an existing image, or create a pair of regions synthetically, for example by drawing two simple polygons. Given the pair of regions, we compare their \raid descriptor with the descriptors of the region pairs in all dataset images. We treat the descriptor values as feature vectors and compare them using the $L_1$ distance, which does not overly penalize single bins that have a high mismatch. In our experiments, we treat all target regions with the same label in an image as a single region. This also improves the robustness of the query, since the segmentation of an image into regions is often ambiguous (e.g. sometimes books in a shelf are annotated individually, sometimes a whole row is annotated as a single region) and regions might be subdivided by occluding objects. We can optionally filter a query by the label of the source or target region. For example, we can query for relationships where the source region has the label `person'. The descriptor for a pair of query regions can also be stored and associated with a specific verb such as `riding' or `surrounding'. This allows future queries to be formulated as sentences consisting of a subject (the label of the source region), a verb (the stored descriptor) and an object (the label of the target region), such as `chairs surrounding table' or `person riding $X$', where $X$ stands for any label. Since \raid is scale-invariant, results may contain relationships between small regions in the background. To filter out these less salient results, we remove source regions with an area below $1\%$ of the image area from the result.

Results of six queries are shown in Figures~\ref{fig:queryresultsA} and~\ref{fig:queryresultsB}. The queries in Figure~\ref{fig:queryresultsA}, as well as the first query in Figure~\ref{fig:queryresultsB} use images from the dataset as query regions. In these queries, we only search for source regions with the label `person'. The remaining two queries use synthetic query regions and search for source and target regions of any label. In the bottom row of each figure, we provide the precision of the first $n$ results of the query as a function of $n$. The ground truth was created by three persons who manually annotated the results of the query in randomized order, without knowledge of our descriptor and without knowledge which method generated the results.
In the `riding' query (Figure~\ref{fig:queryresultsA}, first row), the source region contains an interesting distribution of simple relationship, including source points above and source points in between the target region. Our descriptor successfully finds regions with a similar distribution of simple relationships, while Shape Contexts also return many false positives with a different distribution of simple relationships. Similar results can be observed on the `carrying', `standing on' and `holding' relationships. Note how a similar distribution of simple relationships also corresponds to regions that are intuitively similar to the query. For the two synthetic queries, our method also returns more relevant results. In the `surrounding' query, for example, our descriptor successfully reproduces the gap between source and target region, while Shape Contexts ignore the gap.

\begin{figure*}[t]
	\includegraphics[width=1.0\textwidth]{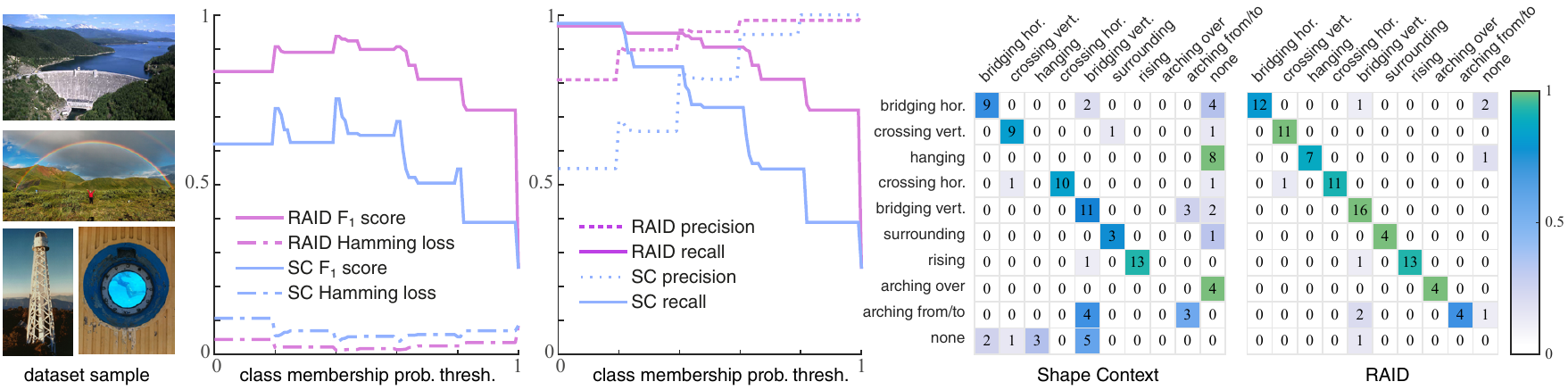}
	\caption{\footnotesize Classification performance on the web image dataset and comparison to Shape Contexts. Four images of the dataset are shown on the left, each contains at least one of the relationship classes. In the center we show various performance measures at different class membership probability thresholds of the binary k-NN classifiers. On the right, the confusion matrices for Shape Contexts and \raid are shown (rows correspond to actual classes, columns to predicted classes, colors are normalized by class size, while numbers show absolute values). Similar to the synthetic dataset, Shape Contexts have a lower performance and have problems detecting some of the classes.}
	\label{fig:webeval}
	\vspace{-12pt}
\end{figure*}

\paragraph{Classification Performance on the Synthetic Dataset} We performed additional evaluation on a small synthetic dataset containing 164 manually created images. Each image shows a single source- and a single target region. These region pairs were labeled manually with zero, one, or multiple labels from among the seven complex relationship classes shown in Figure~\ref{fig:relationships} plus the `surrounded' relationship. Of the 164 relationships, 97 are labeled with one or more relationship classes; the remaining relationships do not correspond to any of the classes. Since relationships can be part of multiple classes (e.g. a bridge may be arching between and bridging two shores), we use multi-label classification. More specifically, we split the multi-label classification into several independent binary classifications, one for each relationship class. Each binary classification is performed by a $k$-NN classifier based on the $L_1$ distance of the \raid descriptors. We set $k=5$, so that the five closest relationships are used to determine the labels of a given relationship.

Results of a leave-one-out cross-validation of the classifier and a comparison to Shape Contexts are shown in Figure~\ref{fig:artificaleval}. Since Shape Contexts only capture simple relationships between a point and a region, they perform poorly for more complex relationships. Note, for example, the large number of relationships that were incorrectly classified as not corresponding to any class, shown in the last column of the confusion matrix. The \raid descriptor captures the \emph{distribution} of simple relationships over a region, resulting in a more discriminative classifier.

\paragraph{Classification Performance on the Web Dataset} The web dataset consists of $69$ images containing a total of $121$ manually labeled relationships. These relationships represent a reasonably balanced mix of the complex relationship classes shown in Figure~\ref{fig:relationships}. Since good examples of a `leaning' relationship are quite uncommon, we used the `rising' relationship (`hanging' mirrored horizontally) instead. Similar to the synthetic dataset, we used one binary $k$-NN classifier with $k=5$ for each relationship class to perform the classification.

Results of a leave-one-out cross-validation and a comparison to Shape Contexts are shown in Figure \ref{fig:webeval}. Note how the results are similar to those of the synthetic dataset. Some classes like `hanging' and `arching over' cannot be detected and many relationships were incorrectly classified as not belonging to any class (last column of the confusion matrix). Our \raid descriptor achieves roughly a $40\%$ increase in the $F_1$ score compared to Shape Contexts and can successfully classify most of the regions.

\begin{figure}[t]
	\includegraphics[width=1.0\columnwidth]{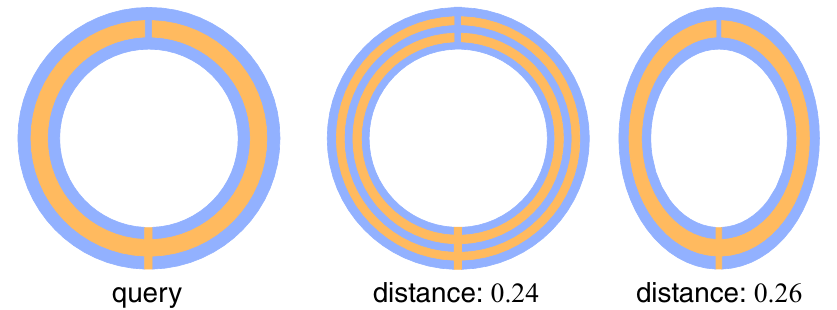}
	\caption{\footnotesize The resolution of our descriptor limits the complexity of relationships that can be captured. The query (left) is more similar to the image in the center than to a deformed version of the query (right), since details in the center are too fine to be described properly. Shown are the $L_1$ distances of the \raid{}s (maximum possible distance is $2$).}
	\label{fig:limitations}
	\vspace{-14pt}
\end{figure}

\paragraph{Limitations} Due to the limited number of bins of our descriptor ($256$ in our experiments), there is a limit to the complexity of the relationships that can be described. An `interleaved' relationship, for example, might be difficult to describe. See Figure~\ref{fig:limitations} for an example. Here, the interleaved rings of the center regions cannot be distinguished properly from rings of the query, since the detail is too fine to be captured by the descriptor bins. Increasing the resolution of the descriptor relieves the problem but also makes the descriptor less tolerant to geometric differences in the relationships. In future work, we would like to experiment with different distance measures, such as the Earth-Movers distance~\cite{Rubner:1998}, which might help to increase the resolution of the descriptor without decreasing the tolerance.

\section{Conclusion}

We have presented \raid, a descriptor for complex relationships between image regions. The key idea of the descriptor is to capture the spatial distribution of simple point-to-region relationship to describe more complex relationships between a pair of regions. To the best of our knowledge, there is currently no descriptor that attempts to capture complex relationships between image regions. Our descriptor is conceptually simple, easy to implement and experiments have shown that it can be employed successfully for relationship-based image retrieval in large databases and for relationship classification, with a clear advantage over Shape Contexts, a descriptor for simple point-to-region relationships.

Continuing this line of research, we would like to extend \raid to describe relationships between 3D models %(similar to the extension of Shape Contexts to 3D models~\cite{Pajdla:2004:3DS}), 
(either given as voxels or polygon meshes),
% experiment with different distance measures for our descriptor, such as the Earth-Movers distance~\cite{Rubner:1998},
use our descriptor in more advanced machine learning techniques, for instance to refine a query by interactively marking good and bad results, and use \raid as a basis to describe the composition of an image, for example by constructing a graph of pair-wise region relationships.
%extend the descriptor to handle ternary or higher-order relationships, such as the `separating' relationship between two different regions (e.g. mountains separate sea and sky).

%\section*{Acknowledgements}

\bibliographystyle{acmsiggraph}
\bibliography{imgrelations}
\end{document}

%% file: abstract.tex
\begin{abstract}
%We, as humans,
As humans, we regularly interpret images based on the relations between image regions. For example, a person {\em riding} object X, or a plank {\em bridging} two objects.
Current methods provide limited support to search for images based on such relations.
%However, limited computational support exists to search for images based on such relations. 
%
We present \raid, a relation-augmented image descriptor that supports queries based on inter-region relations. 
The key idea of our descriptor is to capture the spatial distribution of simple point-to-region relationships to describe more complex relationships between two image regions. 
We evaluate the proposed descriptor by querying into a large subset of the Microsoft COCO database and successfully extract non-trivial images demonstrating complex inter-region relations, which are easily missed or erroneously classified by existing methods. 

\end{abstract}

%% file: introduction.tex
\section{Introduction}

Content-based image retrieval is an important task for image processing applications. For example, an artist may search for a particular scene configuration for inspiration, or a media creator might seek images with a particular assembly of objects. Text-based search, using keywords or tags, is still the most commonly available search option. Advanced alternatives exist that exploit color histograms~\cite{Pentland:1996,Smeulders:2000}, object sketches~\cite{Eitz:2009,YangCao:2011}, or even a rough composition guidance~\cite{Hu2013}.

The last decades have witnessed significant advances in semi-automated and automated image segmentation algorithms. They have resulted in large image databases containing many thousands of labeled and segmented images (e.g., Microsoft COCO~\cite{Lin2014}, Pascal VOC~\cite{Mottaghi:2014}, MIT SUN~\cite{Xiao:2010}). Hence, it is now possible to search for images having regions labeled `horse,' or  `man,' or both `horse' {\em and} `man.' However, there is little support to query based on how the segments are {\em related}. For example, how can we search for images showing `man {\em riding} a horse,' or `man {\em standing next} to a horse,' or more generally `man {\em riding} any object.' 

In this paper, we present \raid as a relation-augmented image descriptor that supports queries based on inter-segment relations.
We identify a set of commonly occurring relations, particularly complex relations (e.g., bridging, riding, leaning, etc.) beyond usual relations like above, below, adjacent, etc. This essentially allows us to query by verbs relating image segment names
%(i.e., nouns),
%(e.g., `person riding horse'),
by associating a particular descriptor with each of the verbs.
Our framework is general in the sense that the user can alternatively sketch a
%description of designed
composition of image segments, or pick a pair of regions in an existing image, and the system can construct an appropriate \raid. An example is given in Figure~\ref{fig:teaser}, where the riding relationship in an existing image is used to query a large annotated image database.
These relationships describe the spatial composition of regions in an image. 
It is important to note that we are tackling a purely two-dimensional problem. Describing the relationship between the actual three-dimensional objects that are represented by the regions is a different problem.

Inter-region relationship are useful in several active research areas, including
%Areas of application include
image editing, image synthesis and content-based image retrieval. They could be used to guide edit propagation~\cite{Berthouzoz:2011:FCP,Yucer2012} by constraining edits to have a given relationship to the edited region,
%propagating edits to regions with similar relationships),
improve library-driven image synthesis~\cite{Hu2013} by returning more relevant regions from the library, and enhance image completion~\cite{Hays:2007:SCU,Huang:2013:MGT} in context-dependent image regions.

Current shape descriptors such as Shape Contexts~\cite{Belongie:2002} are able to describe the relationship between a point and a region, such as `below' or `adjacent.' In a complex relationship between two regions, these simple point-to-region relationships usually vary over a region. Take for example the image in Figure~\ref{fig:descriptoronreal}: the head of the man is above the bench, while his feet are below. The key idea of our descriptor is to capture the spatial distribution of these simple point-to-region relationships to describe more complex relationships between two image regions.
%\raid can be used to search for a specific spatial composition of two image regions in a large database (note that this is different from searching for specific relationships between the actual 3D objects shown in the image) or to classify region pairs based on their relationship.

%\raid describes the composition of two image regions with a four-dimensional histogram: relationships between individual points and an image region are described by a two-dimensional histogram of distance and direction 

%Mention that we are looking at the image composition, not at the 3D objects shown in the image

We evaluate the query performance as well as the classification performance of our descriptor and provide a comparison to Shape Contexts as a baseline shape descriptor. The query performance is measured quantitatively as the precision of query results in a large dataset consisting of 10000 images. Classification performance is tested on two smaller datasets, a synthetic dataset containing 164 images and a set of 75 images collected from the web. Results show that our method is able to successfully describe complex relationships with a clear improvement over Shape Contexts.

%% file: imgrelations.bbl
\begin{thebibliography}{\protect\citename{Berthouzoz et~al\mbox{.} }2011}

\bibitem[\protect\citename{Badadapure }2013]{Badadapure2013}
{\sc Badadapure, P.~R.}
\newblock 2013.
\newblock {Content-Based Image Retrieval by Combining Structural and Content
  Based Features}.
\newblock {\em International Journal of Engineering and Advanced Technology 2},
  4, 154--156.

\bibitem[\protect\citename{Belongie et~al\mbox{.} }2002]{Belongie:2002}
{\sc Belongie, S., Malik, J., and Puzicha, J.}
\newblock 2002.
\newblock Shape matching and object recognition using shape contexts.
\newblock {\em Pattern Analysis and Machine Intelligence, IEEE Transactions on
  24}, 4, 509--522.

\bibitem[\protect\citename{Berthouzoz et~al\mbox{.} }2011]{Berthouzoz:2011:FCP}
{\sc Berthouzoz, F., Li, W., Dontcheva, M., and Agrawala, M.}
\newblock 2011.
\newblock A framework for content-adaptive photo manipulation macros:
  Application to face, landscape, and global manipulations.
\newblock {\em ACM Trans. Graph. 30}, 5 (Oct.), 120:1--120:14.

\bibitem[\protect\citename{Bloch }2005]{Bloch2005}
{\sc Bloch, I.}
\newblock 2005.
\newblock {Fuzzy spatial relationships for image processing and interpretation:
  A review}.
\newblock In {\em Image and Vision Computing}, vol.~23, 89--110.

\bibitem[\protect\citename{Boo }2015]{Boostpolygon}
2015.
\newblock Boost polygon, version 1.58.
\newblock www.boost.org.

\bibitem[\protect\citename{Cao et~al\mbox{.} }2011]{YangCao:2011}
{\sc Cao, Y., Wang, C., Zhang, L., and Zhang, L.}
\newblock 2011.
\newblock Edgel index for large-scale sketch-based image search.
\newblock In {\em Proc. of IEEE Conf. on Comp. Vision and Pattern Recognition},
  761--768.

\bibitem[\protect\citename{Celebi and Aslandogan }2005]{Celebi:2005}
{\sc Celebi, M.~E., and Aslandogan, Y.~A.}
\newblock 2005.
\newblock A comparative study of three moment-based shape descriptors.
\newblock In {\em IEEE Proc. of the Internat. Conf. on Information Technology},
  788--793.

\bibitem[\protect\citename{Chandran and Kiran }2003]{Chandran2003}
{\sc Chandran, S., and Kiran, N.}
\newblock 2003.
\newblock {Image retrieval with embedded region relationships}.
\newblock In {\em Proceedings of the 2003 ACM symposium on Applied computing -
  SAC '03}, 760.

\bibitem[\protect\citename{Flusser }1992]{Flusser:1992}
{\sc Flusser, J.}
\newblock 1992.
\newblock Invariant shape description and measure of object similarity.
\newblock In {\em Image Processing and its Applications, 1992., International
  Conference on}, 139--142.

\bibitem[\protect\citename{Goshtasby }1985]{Goshtasby:1985}
{\sc Goshtasby, A.}
\newblock 1985.
\newblock Description and discrimination of planar shapes using shape matrices.
\newblock {\em IEEE PAMI. 7}, 6, 738--743.

\bibitem[\protect\citename{Hays and Efros }2007]{Hays:2007:SCU}
{\sc Hays, J., and Efros, A.~A.}
\newblock 2007.
\newblock Scene completion using millions of photographs.
\newblock {\em ACM Trans. Graph. 26}, 3 (July).

\bibitem[\protect\citename{Hsieh and Hsu }2008]{Hsieh2008}
{\sc Hsieh, S.~M., and Hsu, C.~C.}
\newblock 2008.
\newblock {Retrieval of images by spatial and object similarities}.
\newblock {\em Information Processing and Management 44}, 3, 1214--1233.

\bibitem[\protect\citename{Hu et~al\mbox{.} }2013]{Hu2013}
{\sc Hu, S.-M., Zhang, F.-L., Wang, M., Martin, R.~R., and Wang, J.}
\newblock 2013.
\newblock {PatchNet: A Patch-based Image Representation for Interactive
  Library-driven Image Editing}.
\newblock {\em ACM Transactions on Graphics 32}, 6, 1--12.

\bibitem[\protect\citename{Huang et~al\mbox{.} }2013]{Huang:2013:MGT}
{\sc Huang, H., Yin, K., Gong, M., Lischinski, D., Cohen-Or, D., Ascher, U.,
  and Chen, B.}
\newblock 2013.
\newblock "mind the gap": Tele-registration for structure-driven image
  completion.
\newblock {\em ACM Trans. Graph. 32}, 6 (Nov.), 174:1--174:10.

\bibitem[\protect\citename{Huang et~al\mbox{.} }2014]{Huang2014}
{\sc Huang, S., Wang, W., and Zhang, H.}
\newblock 2014.
\newblock {Retrieving images using saliency detection and graph matching}.
\newblock In {\em 2014 IEEE Int. Conference on Image Processing (ICIP)},
  3087--3091.

\bibitem[\protect\citename{Karpathy and Li }2015]{Karpathy2015}
{\sc Karpathy, A., and Li, F.-F.}
\newblock 2015.
\newblock {Deep Visual-Semantic Alignments for Generating Image Descriptions}.
\newblock In {\em CVPR}.

\bibitem[\protect\citename{Kazmi et~al\mbox{.} }2013]{Kazmi:2013}
{\sc Kazmi, I.~K., You, L., and Zhang, J.~J.}
\newblock 2013.
\newblock A survey of 2d and 3d shape descriptors.
\newblock {\em 2014 11th International Conference on Computer Graphics, Imaging
  and Visualization 0\/}, 1--10.

\bibitem[\protect\citename{Ko and Byun }2002]{Ko2002}
{\sc Ko, B., and Byun, H.}
\newblock 2002.
\newblock {Multiple Regions and Their Spatial Relationship-Based Image
  Retrieval}.
\newblock In {\em LNCS 2383}. 81--90.

\bibitem[\protect\citename{Kulkarni et~al\mbox{.} }2013]{Kulkarni2013}
{\sc Kulkarni, G., Premraj, V., Ordonez, V., Dhar, S., Li, S., Choi, Y., Berg,
  A.~C., and Berg, T.~L.}
\newblock 2013.
\newblock {Baby talk: Understanding and generating simple image descriptions}.
\newblock {\em IEEE Trans. on Pattern Anal. and Mach. Intell. 35}, 12,
  2891--2903.

\bibitem[\protect\citename{Lan et~al\mbox{.} }2012]{Lan2012}
{\sc Lan, T., Yang, W., Wang, Y., and Mori, G.}
\newblock 2012.
\newblock {Image retrieval with structured object queries using latent ranking
  SVM}.
\newblock In {\em Lect. Notes in Computer Science}, vol.~7577 LNCS, 129--142.

\bibitem[\protect\citename{Lee and Hwang }2002]{Lee2002}
{\sc Lee, S. L.~S., and Hwang, E. H.~E.}
\newblock 2002.
\newblock {Spatial similarity and annotation-based image retrieval system}.
\newblock {\em Proceedings of Fourth Int. Symposium on Multimedia Software
  Engineering\/}.

\bibitem[\protect\citename{Lin et~al\mbox{.} }2014]{Lin2014}
{\sc Lin, T., Maire, M., Belongie, S., Hays, J., Perona, P., Ramanan, D.,
  Doll{\'{a}}r, P., and Zitnick, C.~L.}
\newblock 2014.
\newblock Microsoft {COCO:} common objects in context.
\newblock {\em CoRR abs/1405.0312\/}.

\bibitem[\protect\citename{Lowe }2004]{Lowe04}
{\sc Lowe, D.}
\newblock 2004.
\newblock Distinctive image features from scale-invariant keypoints.
\newblock {\em Int. Journal of Computer Vision 60}, 2, 91--110.

\bibitem[\protect\citename{Malisiewicz and Efros }2009]{Malisiewicz2009}
{\sc Malisiewicz, T., and Efros, A.~A.}
\newblock 2009.
\newblock {Beyond Categories: The Visual Memex Model for Reasoning About Object
  Relationships}.
\newblock In {\em NIPS}, 1--9.

\bibitem[\protect\citename{Mathias~Eitz and Alexa }2009]{Eitz:2009}
{\sc Mathias~Eitz, Kristian~Hildebrand, T.~B., and Alexa, M.}
\newblock 2009.
\newblock A descriptor for large scale image retrieval based on sketched
  feature lines.
\newblock In {\em Eurographics Symposium on Sketch-Based Interfaces and
  Modeling}, 29--38.

\bibitem[\protect\citename{Mottaghi et~al\mbox{.} }2014]{Mottaghi:2014}
{\sc Mottaghi, R., Chen, X., Liu, X., Cho, N.-G., Lee, S.-W., Fidler, S.,
  Urtasun, R., and Yuille, A.}
\newblock 2014.
\newblock The role of context for object detection and semantic segmentation in
  the wild.
\newblock In {\em IEEE CVPR}.

\bibitem[\protect\citename{Pentland et~al\mbox{.} }1996]{Pentland:1996}
{\sc Pentland, A., Picard, R.~W., and Sclaroff, S.}
\newblock 1996.
\newblock Photobook: Content-based manipulation of image databases.
\newblock {\em Int. J. Comput. Vision 18}, 3 (June), 233--254.

\bibitem[\protect\citename{Rubner et~al\mbox{.} }1998]{Rubner:1998}
{\sc Rubner, Y., Tomasi, C., and Guibas, L.~J.}
\newblock 1998.
\newblock A metric for distributions with applications to image databases.
\newblock In {\em Proc. of the Sixth International Conference on Computer
  Vision}, IEEE Computer Society, Washington, DC, USA, ICCV '98, 59--66.

\bibitem[\protect\citename{Smeulders et~al\mbox{.} }2000]{Smeulders:2000}
{\sc Smeulders, A. W.~M., Worring, M., Santini, S., Gupta, A., and Jain, R.}
\newblock 2000.
\newblock Content-based image retrieval at the end of the early years.
\newblock {\em IEEE Trans. Pattern Anal. Mach. Intell. 22}, 12 (Dec.),
  1349--1380.

\bibitem[\protect\citename{Teague }1980]{Teague:80}
{\sc Teague, M.~R.}
\newblock 1980.
\newblock Image analysis via the general theory of moments$\ast$.
\newblock {\em J. Opt. Soc. Am. 70}, 8 (Aug), 920--930.

\bibitem[\protect\citename{Wang }2003]{Wang2003}
{\sc Wang, Y.-H.}, 2003.
\newblock {Image indexing and similarity retrieval based on spatial
  relationship model}.

\bibitem[\protect\citename{Xiao et~al\mbox{.} }2010]{Xiao:2010}
{\sc Xiao, J., Hays, J., Ehinger, K., Oliva, A., and Torralba, A.}
\newblock 2010.
\newblock Sun database: Large-scale scene recognition from abbey to zoo.
\newblock In {\em IEEE CVPR}.

\bibitem[\protect\citename{Y\"{u}cer et~al\mbox{.} }2012]{Yucer2012}
{\sc Y\"{u}cer, K., Jacobson, A., Hornung, A., and Sorkine, O.}
\newblock 2012.
\newblock Transfusive image manipulation.
\newblock {\em ACM Trans. Graph. 31}, 6 (Nov.), 176:1--176:9.

\bibitem[\protect\citename{Zhang and Lu }2004]{Zhang:2004}
{\sc Zhang, D., and Lu, G.}
\newblock 2004.
\newblock Review of shape representation and description techniques.
\newblock {\em Pattern Recognition 37}, 1, 1 -- 19.

\bibitem[\protect\citename{Zhou et~al\mbox{.} }2001]{Zhou2001}
{\sc Zhou, X.~M., Ang, C.~H., and Ling, T.~W.}
\newblock 2001.
\newblock {Image retrieval based on object's orientation spatial relationship}.
\newblock {\em Pattern Recognition Letters 22}, 5, 469--477.

\end{thebibliography}
